\documentclass[aps,prd,twocolumn,preprintnumbers,
superscriptaddress,showpacs,floatfix]{revtex4}

\usepackage{bm}
\usepackage{epsfig}
\usepackage{graphics}

\begin{document}

\preprint{FISIST/22-2002/CFIF}
\preprint{UT-STPD-4/02}
\preprint{SISSA-74/02}

\title{On Yukawa quasi-unification
with ${\bm\mu\bm<\bm0}$}

\author{M.E. G\'{o}mez}
\email{mgomez@cfif.ist.utl.pt}
\affiliation{Centro de F\'{\i}sica das
Interac\c{c}\~{o}es Fundamentais (CFIF),
Departamento de F\'{\i}sica, Instituto Superior
T\'{e}cnico, Av. Rovisco Pais,
Lisboa 1049-001, Portugal}
\author{G. Lazarides}
\email{lazaride@eng.auth.gr}
\affiliation{Physics Division, School of Technology,
Aristotle University of Thessaloniki,
Thessaloniki 54124, Greece}
\author{C. Pallis}
\email{pallis@sissa.it}
\affiliation{Scuola Internazionale
Superiore Di Studi Avanzati (SISSA),
Via Beirut 2-4, Trieste 34013, Italy}

\date{\today}

\begin{abstract}
Although recent data on the muon anomalous
magnetic moment strongly disfavor the
constrained minimal supersymmetric standard
model with $\mu<0$, they cannot exclude it
because of theoretical ambiguities. We
consider this model supplemented by a
Yukawa quasi-unification condition which
allows an acceptable $b$-quark mass. We
find that the cosmological upper bound on
the lightest sparticle relic abundance is
incompatible with the data on the branching
ratio of $b\rightarrow s\gamma$, which is
evaluated by including all the
next-to-leading order corrections. Thus,
this scheme is not viable.

\end{abstract}

\maketitle

\par
The constrained minimal supersymmetric standard
model (CMSSM), which is a predictive
version of the minimal supersymmetric standard
model (MSSM) based on universal boundary
conditions, can be further restricted by being
embedded in a supersymmetric (SUSY) grand
unified theory (GUT) with a gauge group
containing $SU(4)_c$ and $SU(2)_R$. This can
lead \cite{pana} to Yukawa unification
\cite{als}, i.e., the exact unification of the
third generation `asymptotic' Yukawa couplings.
However, for both signs of the MSSM parameter
$\mu$, the CMSSM supplemented by the
assumption of exact Yukawa unification yields
unacceptable values of the $b$-quark mass.
This is due to the generation of sizeable SUSY
corrections to $m_b$ \cite{copw} which have
the sign of $\mu$. The predicted tree-level
$m_b(M_Z)$, which turns out to be close to
the upper edge of its $95\%$ confidence level
(c.l.) experimental range $2.684-3.092~{\rm GeV}$,
receives, for $\mu>0$ ($\mu<0$), large positive
(negative) corrections which drive it well above
(a little below) the allowed range. This range is
derived from the $95\%$ c.l. range for $m_b(m_b)$
in the $\overline{MS}$ scheme $3.95-4.55~{\rm GeV}$
\cite{mb} evolved up to $M_Z$ in accord with the
analysis of Ref.~\cite{baermb} and with
$\alpha_s(M_Z)=0.1185$.

\par
In Ref.~\cite{qcdm}, we have built SUSY GUT
models based on the Pati-Salam gauge group
$SU(4)_c\times SU(2)_L\times SU(2)_R$ which
moderately violate Yukawa unification and,
thus, can allow an acceptable $b$-quark mass
for both signs of $\mu$ even with universal
boundary conditions. These models yield a
class of `asymptotic' Yukawa
quasi-unification conditions (YQUCs)
replacing exact Yukawa unification. Supplementing
the CMSSM with one of these YQUCs, we achieved
\cite{qcdm}, for $\mu>0$, an acceptable
$m_b(M_Z)$ in accord with all the
phenomenological and cosmological requirements.
The corresponding SUSY GUT model
leads \cite{nshi} to a new version of shifted
hybrid inflation \cite{jean}.

\par
Here, we will examine whether the CMSSM with
a YQUC can yield an acceptable $m_b(M_Z)$
in accord with all the available constraints
for $\mu<0$ too. In this case, compatibility
of the $b$-quark mass with the data requires
a deviation from Yukawa unification which is
much smaller than the one needed for $\mu>0$.
Consequently, there is no need to extend the
simplest SUSY GUT model of Ref.~\cite{qcdm},
which yields a suppressed violation of
Yukawa unification originating from
non-renormalizable interactions.

\par
The CMSSM with $\mu<0$ is now strongly
disfavored (but not excluded \cite{martin})
by the constraint on the deviation, $\delta
a_\mu$, of the measured value of the muon
anomalous magnetic moment $a_\mu$ from its
predicted value in the standard model (SM)
$a^{\rm SM}_\mu$. The recent data \cite{nmuon}
on $a_\mu$ confirm the earlier ones
\cite{muon} but their precision is twice that
of the previous data. On the other hand,
$a^{\rm SM}_\mu$ is not yet stabilized. This
is mainly due to
the hadronic vacuum polarization contribution.
According to the most updated \cite{davier}
evaluation of this contribution, there is
a considerable discrepancy between the
results based on $e^+e^-$ data and on $\tau$
data. Taking into account both results, we
obtain the $95\%$ c.l. range $-4.7\times
10^{-10}\lesssim\delta a_\mu\lesssim 56\times
10^{-10}$, where the negative values are
allowed only by the $\tau$-based calculation.
Using the formulae of Ref.~\cite{gmuon}, we
calculate $\delta a_\mu$ in the CMSSM and we
observe that the result is negative (positive)
for $\mu<0$ ($\mu>0)$ and increases (decreases)
as the lightest sparticle (LSP) mass
$m_{\rm LSP}$ increases. Consequently, a lower
bound on $m_{\rm LSP}$ can be derived from the
lower (upper) experimental limit on
$\delta a_\mu$ for $\mu<0$ ($\mu>0)$. Also, we
see that $\mu<0$ is allowed only by the
$\tau$-based calculation.

\par
The CMSSM with $\mu<0$ is, also, severely
restricted by the recent experimental results
\cite{cleo} on the inclusive branching ratio
${\rm BR}(b\rightarrow s\gamma)$. We compute
this branching ratio by using an updated
version of the relevant calculation in the
{\tt micrOMEGAs} code \cite{micro}. Comparing
carefully this version with our calculation
of ${\rm BR}(b\rightarrow s\gamma)$ in
Ref.~\cite{qcdm}, we concluded that this
improved version is complete and reliable.
The SM contribution is calculated by using
the formalism of Ref.~\cite{kagan} and
incorporating the improvements of
Ref.~\cite{gambino}. The charged Higgs boson
($H^\pm$) contribution is evaluated by
including the next-to-leading order (NLO)
QCD corrections from Ref.~\cite{nlohiggs}
and the $\tan\beta$ enhanced contributions
from Ref.~\cite{nlosusy}. The dominant SUSY
contribution includes resummed NLO SUSY QCD
corrections from Ref.~\cite{nlosusy}, which
hold for large $\tan\beta$. The $H^\pm$
contribution interferes constructively with
the SM contribution, while the SUSY one
interferes constructively (destructively)
with the other two contributions when
$\mu<0$ ($\mu>0)$. The SM plus $H^\pm$ and
the SUSY contributions decrease as
$m_{\rm LSP}$ increases and so, for $\mu<0$,
an additional lower bound on $m_{\rm LSP}$
can be derived \cite{borzumati} from the
upper experimental limit on this branching
ratio. At $95\%$ c.l., the experimental range
of ${\rm BR}(b\rightarrow s\gamma)$ is
estimated \cite{qcdm} by combining
appropriately the various experimental and
theoretical errors involved. The result
is $1.9\times 10^{-4}\lesssim {\rm BR}
(b\rightarrow s\gamma)\lesssim4.6 \times
10^{-4}$. We find that the NLO corrections
reduce ${\rm BR}(b\rightarrow s\gamma)$
by $10-15\%$. In
Ref.~\cite{cd2} (which adopts the opposite
sign convention for $\mu$), where the
$\tan\beta$ enhanced and NLO SUSY QCD
corrections were ignored, the reduction was
considerably higher yielding a less stringent
restriction.

\par
\begin{figure}[t]
\includegraphics[width=60mm,angle=-90]{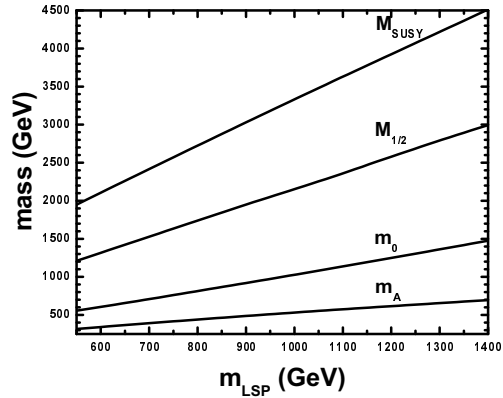}
\caption{\label{Mnx} The mass parameters $m_A$,
$m_0$, $M_{1/2}$ and $M_{\rm SUSY}$ versus
$m_{\rm LSP}$ for $A_0=0,~\Delta_{\tilde\tau_2}=0$
and $m_b(M_Z)=2.888~{\rm GeV}$.}
\end{figure}

The LSP mass is bounded above by the
requirement that the LSP relic abundance
$\Omega_{\rm LSP}h^2$ in the universe does not
exceed the upper limit on the cold dark
matter (CDM) abundance derived from observations.
Recent results \cite{dasi} suggest that, at
$95\%$ c.l., $\Omega_{\rm LSP}h^2\lesssim 0.22$.
In general, $\Omega_{\rm LSP}h^2$ increases with
$m_{\rm LSP}$. So, its value is expected to be
quite large due to the large $m_{\rm LSP}$'s
required by the constraints from $\delta a_\mu$
and, especially, $b\rightarrow s\gamma$. The
$\Omega_{\rm LSP}h^2$ is calculated by using
{\tt micrOMEGAs} \cite{micro}, which is the most
complete code available at present. It includes
accurately thermally averaged exact tree-level
cross sections of all possible (co)annihilation
processes and treats pole effects properly with
one-loop QCD corrected decay widths. In the
model which we will consider, the LSP is an
almost pure bino ($\tilde\chi$) and the
next-to-LSP (NLSP) is the lightest stau
($\tilde\tau_2$). Important annihilation
channels are not only the ones with fermions
$f\bar f$ in the final state, but also the
ones with $HZ$, $W^\pm H^\mp$, $hA$ \cite{nra}
($h$, $H$ are the CP-even neutral Higgs bosons
and $A$ the CP-odd neutral Higgs boson).~On the
other hand, as we can observe from the sparticle
spectrum of the model (see Fig.~\ref{Mnx}),
$m_A$ is much smaller than $2m_{\rm LSP}$. So,
the LSP annihilation via the s-channel exchange
of an $A$-boson is not enhanced as in the
$\mu>0$ case \cite{qcdm, cmssm}. As a
consequence, the only available mechanism for
reducing $\Omega_{\rm LSP}h^2$ is bino-stau
coannihilation \cite{qcdm,ellis,cdm,cd2} (for
an updated analysis, see Ref.~\cite{roberto}),
which is activated when the relative mass
splitting between the $\tilde\tau_2$ and the
LSP, $\Delta_{\tilde\tau_2}=(m_{\tilde\tau_2}-
m_{\rm LSP})/m_{\rm LSP}$, is reduced below
about 0.25. For fixed $m_{\rm LSP}$,
$\Omega_{\rm LSP}h^2$ decreases with
$\Delta_{\tilde\tau_2}$, since coannihilation
becomes more efficient. So the CDM bound
yields an upper limit on $\Delta_{\tilde\tau_2}$.

We will consider here the CMSSM with $\mu<0$
which results from the simplest SUSY GUT model
constructed in Ref.~\cite{qcdm} without the
extra Higgs superfields $\bar{\phi}$, $\phi$,
$\bar{\phi}^{\prime}$, $\phi^{\prime}$. To
reduce the number of free parameters, we will
further assume that the non-renormalizable
$SU(2)_R$ triplet coupling of Ref.~\cite{qcdm}
which leads to violation of Yukawa unification
is suppressed. The violation is then achieved
predominantly via the non-renormalizable
$SU(2)_R$ singlet coupling of Ref.~\cite{qcdm}
and, thus, $\alpha_1=\alpha_2$ in Eq.~(15) of
this reference, with $\alpha_1$, $\alpha_2$ of
order 0.1 or smaller. Note that the other
alternative with the triplet coupling
dominating, which was studied in
Ref.~\cite{qcdm} with $\mu>0$, is not viable
for $\mu<0$ since it leads to a charged LSP.
Under the above assumptions, the YQUC takes
the form:
\begin{equation}
h_t:h_b:h_\tau=|1-c|:|1-c|:|1+3c|,
\label{minimal}
\end{equation}
where $c$ is a complex parameter of order 0.1
or smaller. We will take $c<0$, where the
maximal splitting between $h_b$, $h_\tau$ and,
thus, the largest enhancement of $m_b(M_Z)$
is achieved for fixed $|c|\neq 0$. For
$-1/3<c<0$, this splitting is $-\delta h
\equiv(h_b-h_\tau)/h_b=-4c/(1-c)$. We see that
the tau Yukawa coupling becomes smaller than
the top and bottom Yukawa couplings which
remain unified, in contrast to the $\mu>0$
case of Ref.~\cite{qcdm}, where the tau and
bottom Yukawa couplings split from the top
Yukawa coupling by the same amount. The SUSY
GUT model used leads \cite{jean} to successful
shifted hybrid inflation.

\par
Below the GUT scale $M_{\rm GUT}$, this SUSY
GUT model reduces to the CMSSM with universal
soft SUSY breaking scalar masses $m_0$,
gaugino masses $M_{1/2}$, and trilinear scalar
couplings $A_0$ at $M_{\rm GUT}$. It also
implies the YQUC in Eq.~(\ref{minimal}). We
closely follow the renormalization group and
radiative electroweak breaking analysis of
Ref.~\cite{qcdm}. The conditions for
electroweak breaking are imposed on an
optimized scale $M_{\rm SUSY}=
(m_{\tilde t_1}m_{\tilde t_2})^{1/2}$
($\tilde t_{1,2}$ are the stop mass eigenstates),
where the sparticle spectrum is also evaluated.
We incorporate the two-loop corrections to the
CP-even neutral Higgs boson mass matrix by using
{\tt FeynHiggsFast} \cite{fh} and the SUSY
corrections to $m_b(M_{\rm SUSY})$ and
$m_\tau(M_{\rm SUSY})$ from Ref.~\cite{pierce}.
The corrections to $m_\tau(M_{\rm SUSY})$ lead
\cite{cd2} to a small enhancement of
$\tan\beta$. For $m_t(m_t)=166~{\rm GeV}$,
$m_\tau(M_Z)=1.746~{\rm GeV}$, and any
given $m_b(M_Z)$ in its $95\%$ c.l.
range ($2.684-3.092~{\rm GeV}$), we can
determine the parameters $c$ and $\tan\beta$
at $M_{\rm SUSY}$ so that the YQUC in
Eq.~(\ref{minimal}) is satisfied. We are then
left with only three free input parameters
$m_0$, $M_{1/2}$ and $A_0$. The first two
can be replaced \cite{cd2} by $m_{\rm LSP}$
and $\Delta_{\tilde\tau_2}$. In Fig.~\ref{Mnx},
we present the values of $m_A$, $m_0$,
$M_{1/2}$ and $M_{\rm SUSY}$ versus $m_{\rm LSP}$
for $A_0=0$, $\Delta_{\tilde\tau_2}=0$,
$m_b(M_Z)=2.888~{\rm GeV}$. These values are
affected very little by varying $m_b(M_Z)$.
Our scheme, unlike the $\mu>0$ model of
Ref.~\cite{qcdm}, yields $2m_{\rm LSP}\gg m_A$.

\par
\begin{figure}[t]
\includegraphics[width=60mm,angle=-90]{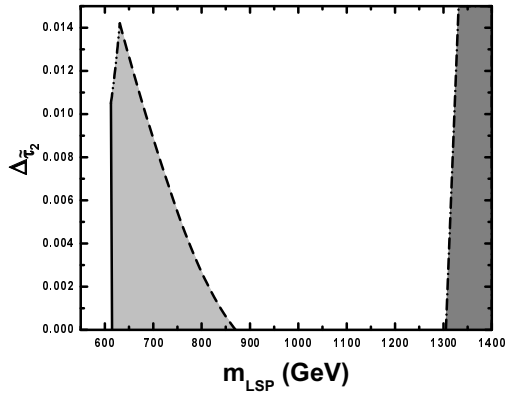}
\caption{\label{A0x} Restrictions on the
$m_{\rm LSP}-\Delta_{\tilde\tau_2}$ plane for
$A_0=0$ and $2.684~{\rm GeV}\lesssim m_b(M_Z)
\lesssim3.092~{\rm GeV}$. The solid
(dot-dashed) line is the lower bound on
$m_{\rm LSP}$ from $\delta a_\mu\gtrsim
-4.7\times 10^{-10}$ (${\rm BR}(b\rightarrow
s\gamma)\lesssim4.6\times 10^{-4}$). The
dashed (double dot-dashed) line is the
upper bound on $m_{\rm LSP}$
($\Delta_{\tilde\tau_2}$) from
$\Omega_{\rm LSP}h^2\lesssim 0.22$.}
\end{figure}

The restrictions on the $m_{\rm LSP}-
\Delta_{\tilde\tau_2}$ plane for $A_0=0$ and
any $m_b(M_Z)$ in its $95\%$ c.l. range are
given in Fig.~\ref{A0x}. The lower bound
on $m_{\rm LSP}$ from $\delta a_\mu\gtrsim
-4.7\times 10^{-10}$ corresponds to $m_b(M_Z)
\simeq 3.092~{\rm GeV}$ and is represented
by a solid line. The maximal
$\Delta_{\tilde\tau_2}$ on this line is
about $0.0105$ and yields $\Omega_{\rm LSP}
h^2\simeq 0.22$ for the same value of
$m_b(M_Z)$. As $m_b(M_Z)$ decreases, the
maximal $\Delta_{\tilde\tau_2}$ from
$\Omega_{\rm LSP}h^2\lesssim 0.22$ increases
along the double dot-dashed line and
reaches its overall maximal value
$\Delta_{\tilde\tau_2}\simeq 0.0142$ at
$m_b(M_Z)\simeq 2.684~{\rm GeV}$. The upper
bound on $m_{\rm LSP}$ from $\Omega_{\rm LSP}
h^2\lesssim 0.22$ is achieved at $m_b(M_Z)
\simeq 2.684~{\rm GeV}$ and corresponds to
the dashed line.
In the lightly shaded region allowed by
$\delta a_\mu$ and CDM considerations, $612.4
~{\rm GeV}\lesssim m_{\rm LSP}\lesssim 873.4
~{\rm GeV}$. The maximal $m_{\rm LSP}$ is
achieved at $\Delta_{\tilde\tau_2}=0$ yielding
${\rm BR}(b\rightarrow s\gamma)\simeq 5.26\times
10^{-4}$. The lower bound on $m_{\rm LSP}$
(dot-dashed line) from ${\rm BR}(b\rightarrow
s\gamma)\lesssim 4.6\times 10^{-4}$ corresponds
to $m_b(M_Z)\simeq 3.092~{\rm GeV}$. In the
corresponding allowed (dark shaded) area,
$m_{\rm LSP}\gtrsim 1305.04~{\rm GeV}$ with
the minimal $m_{\rm LSP}$ achieved at
$\Delta_{\tilde\tau_2}=0$ and yielding
$\Omega_{\rm LSP}h^2\simeq 0.65$. So,
for $A_0=0$, there is no region where all the
restrictions are satisfied. Note that the
constraints ${\rm BR}(b\rightarrow s\gamma)
\gtrsim1.9\times 10^{-4}$ and $\delta a_\mu
\lesssim56\times 10^{-10}$ always hold for the
CMSSM with $\mu<0$. Also, $m_h\gtrsim 114.4
~{\rm GeV}$ \cite{higgs} is valid due to the
heavy SUSY spectrum.

\begin{figure}[t]
\includegraphics[width=60mm,angle=-90]{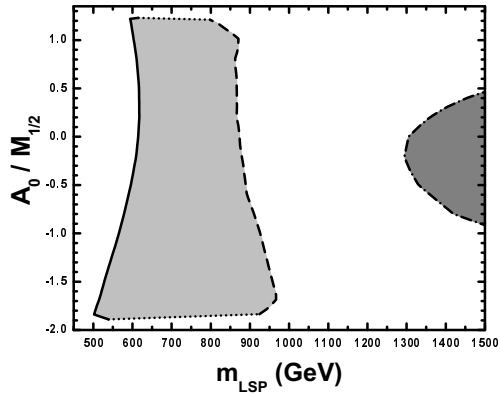}
\caption{\label{Anz} Restrictions on the
$m_{\rm LSP}-A_0/M_{1/2}$ plane for any
$\Delta_{\tilde\tau_2}$ and
$2.684~{\rm GeV}\lesssim m_b(M_Z)\lesssim
3.092~{\rm GeV}$. The notation is generally
as in Fig.~\ref{A0x}. The dotted line and
the inclined parts of the solid line are
explained in the main text.}
\end{figure}

\par
Departure from the $A_0=0$ case does not
change our conclusion as we can see from
Fig.~\ref{Anz}, where the restrictions
on the $m_{\rm LSP}-A_0/M_{1/2}$ plane are
presented for any $\Delta_{\tilde\tau_2}$
and any allowed $m_b(M_Z)$. Actually, we can
just take $\Delta_{\tilde\tau_2}=0$, where
all the constraints happen to become as
less restrictive as possible for any given
$A_0$ (see Fig.~\ref{A0x}). We follow the
same notation for the various lines and
areas as in Fig.~2. For $-1.84\lesssim
A_0/M_{1/2}\lesssim1.21$ ($-1.67\lesssim
A_0/M_{1/2}\lesssim1.03$), the solid
(dashed) line corresponds to $\delta a_\mu
\simeq -4.7\times 10^{-10}$
($\Omega_{\rm LSP}h^2\simeq 0.22$) with
$m_b(M_Z)\simeq 3.092~{\rm GeV}$
($2.684~{\rm GeV}$). As $|A_0|$ increases,
$m_A$ decreases and reaches its lowest
allowed value ($\simeq 100~{\rm GeV}$ for
the $\tan\beta$'s encountered here
\cite{cdf}) on the boundaries of the
lightly shaded regions
for fixed $A_0$. Allowing $m_b(M_Z)$ to
increase from its lower value, we then
obtain the upper and lower inclined parts
of the dashed line corresponding to
$m_A\simeq 100~{\rm GeV}$. Their end
points lie at $A_0/M_{1/2}\simeq 1.21,
-1.84$, where $m_b(M_Z)$ is maximized.
At the maximal (minimal) $m_{\rm LSP}$ on
the dotted line (inclined parts of the
solid line), $\Omega_{\rm LSP}h^2\lesssim
0.22$ ($\delta a_\mu\gtrsim -4.7\times
10^{-10}$) is not saturated, due to the
combined requirements $m_A\gtrsim 100~
{\rm GeV}$ and $m_{\tilde\tau_2}\gtrsim
m_{\tilde\chi}$.

\par
The maximal $m_{\rm LSP}$ in the lightly
shaded area of Fig.~\ref{Anz}, which is
allowed by $\delta a_\mu$ and CDM
considerations, and for $A_0/M_{1/2}>0$
($<0$) is $m_{\rm LSP}\simeq
868~{\rm GeV}$ ($967.3~{\rm GeV}$) and is
achieved at $A_0/M_{1/2}\simeq 1.03$
($-1.67$). This corresponds to the upper
(lower) corner of the dashed line, where
${\rm BR}(b\rightarrow s\gamma)\simeq 5.58
\times 10^{-4}$ ($5.45\times 10^{-4}$).
Note that, for $A_0/M_{1/2}<0$, the
processes $\tilde\tau_2\tilde\tau_2^\ast
\rightarrow W^\pm H^\mp$, $Z A$ become
more efficient. Their total contribution
to the effective cross section is $3-6.5\%$
as $A_0/M_{1/2}$ decreases from $0$ to
$-1.67$. So, coannihilation is strengthened
and bigger $m_{\rm LSP}$'s are allowed
than for $A_0/M_{1/2}>0$. The minimal
$m_{\rm LSP}$ in the dark shaded area,
allowed by $b\rightarrow s\gamma$, is
$m_{\rm LSP} \simeq 1295~{\rm GeV}$ and is
achieved at $A_0/M_{1/2}\simeq -0.2$, where
$\Omega_{\rm LSP}h^2\simeq 0.62$. Thus, even
for $A_0\neq 0$, there is no range of
parameters allowed by all the constraints.

\par
In the lightly shaded area of
Fig.~\ref{Anz}, $\tan\beta$ ($c$) ranges
between about $44.1$ ($-0.09$) and $51.5$
($-0.007$). These values are achieved at
the lowest corner of this area
($A_0/M_{1/2}\simeq -1.89$, $m_b(M_Z)\simeq
3.092~{\rm GeV}$) and the upper corner of
the dashed line ($A_0/M_{1/2}\simeq 1.03$,
$m_b(M_Z)\simeq 2.684~{\rm GeV}$) respectively.
Consequently, the splitting $-\delta h$
ranges from about $0.33$ to $0.028$. However,
fixing $m_b(M_Z)$ to its central value, the
range of $-\delta h$ is reduced to
$0.13-0.09$ corresponding to $47.6\lesssim
\tan\beta\lesssim49$ and $-0.038\lesssim c
\lesssim-0.024$.

\par
The present investigation is an improved
version of our analysis in Ref.~\cite{cd2}
(the sign of $\mu$ there is opposite to the
one adopted here). The main improvements are
the replacement of Yukawa unification by the
YQUC of Eq.~(\ref{minimal}) in connection
with the more stringent present bounds on
$m_b$, the consideration of the $\delta
\alpha_\mu$ constraint \cite{gmuon}, the
inclusion of the NLO SUSY QCD and
$\tan\beta$ enhanced corrections
\cite{nlosusy} in ${\rm BR}(b\rightarrow
s \gamma)$, and the evaluation of
$\Omega_{\rm LSP}h^2$ by the updated code of
Ref.~\cite{micro}. These improvements turn
out to be of crucial importance overruling
our previous conclusion. Our results are
essentially unaffected by allowing
$\alpha_s(M_Z)$ to vary in its $95\%$ c.l.
range, which slightly enlarges
\cite{qcdm,baermb} the range of $m_b(M_Z)$.
This is due to the fact that, here, the LSP
annihilation to $b\bar b$ via an $A$-pole
exchange is subdominant in
$\Omega_{\rm LSP}h^2$.

\par
In summary, we studied the CMSSM with $\mu<0$
and a YQUC from the simplest SUSY GUT of
Ref.~\cite{qcdm}. We imposed the constraints
from $m_b$, $\delta\alpha_\mu$, $b\rightarrow
s\gamma$ and CDM. Although the NLO corrections
to $b\rightarrow s\gamma$ (coannihilations)
drastically reduce (enhance) the
lower (upper) bound on $m_{\rm LSP}$, the
$b\rightarrow s\gamma$ and CDM requirements
remain incompatible. Thus, despite the fact
that, with the $\tau$-based calculation of
$\alpha^{\rm SM}_\mu$, the $\delta\alpha_\mu$
and CDM criteria can be simultaneously valid,
this model is excluded.

\vspace{0.25cm}

\par
We are grateful to the {\tt micrOMEGAs} team,
G. B\'{e}langer, F. Boudjema, A. Pukhov and
A. Semenov, for providing us with their
updated code for calculating ${\rm BR}
(b\rightarrow s\gamma)$. We also thank J. Ellis,
W. Liao, A. Masiero and E. Pallante  for useful
discussions. This work was supported by European
Union under the RTN contracts HPRN-CT-2000-00148
and HPRN-CT-2000-00152. M.E.G. acknowledges
support from the `Funda\c c\~ao para a Ci\^encia
e Tecnologia' under contract SFRH/BPD/5711/2001.

\def\ijmp#1#2#3{{Int. Jour. Mod. Phys.}
{\bf #1},~#3~(#2)}
\def\plb#1#2#3{{Phys. Lett. B }{\bf #1},~#3~(#2)}
\def\zpc#1#2#3{{Z. Phys. C }{\bf #1},~#3~(#2)}
\def\prl#1#2#3{{Phys. Rev. Lett.}
{\bf #1},~#3~(#2)}
\def\rmp#1#2#3{{Rev. Mod. Phys.}
{\bf #1},~#3~(#2)}
\def\prep#1#2#3{{Phys. Rep. }{\bf #1},~#3~(#2)}
\def\prd#1#2#3{{Phys. Rev. D }{\bf #1},~#3~(#2)}
\def\npb#1#2#3{{Nucl. Phys. }{\bf B#1},~#3~(#2)}
\def\npps#1#2#3{{Nucl. Phys. B (Proc. Sup.)}
{\bf #1},~#3~(#2)}
\def\mpl#1#2#3{{Mod. Phys. Lett.}
{\bf #1},~#3~(#2)}
\def\arnps#1#2#3{{Annu. Rev. Nucl. Part. Sci.}
{\bf #1},~#3~(#2)}
\def\sjnp#1#2#3{{Sov. J. Nucl. Phys.}
{\bf #1},~#3~(#2)}
\def\jetp#1#2#3{{JETP Lett. }{\bf #1},~#3~(#2)}
\def\app#1#2#3{{Acta Phys. Polon.}
{\bf #1},~#3~(#2)}
\def\rnc#1#2#3{{Riv. Nuovo Cim.}
{\bf #1},~#3~(#2)}
\def\ap#1#2#3{{Ann. Phys. }{\bf #1},~#3~(#2)}
\def\ptp#1#2#3{{Prog. Theor. Phys.}
{\bf #1},~#3~(#2)}
\def\apjl#1#2#3{{Astrophys. J. Lett.}
{\bf #1},~#3~(#2)}
\def\n#1#2#3{{Nature }{\bf #1},~#3~(#2)}
\def\apj#1#2#3{{Astrophys. J.}
{\bf #1},~#3~(#2)}
\def\anj#1#2#3{{Astron. J. }{\bf #1},~#3~(#2)}
\def\mnras#1#2#3{{MNRAS }{\bf #1},~#3~(#2)}
\def\grg#1#2#3{{Gen. Rel. Grav.}
{\bf #1},~#3~(#2)}
\def\s#1#2#3{{Science }{\bf #1},~#3~(#2)}
\def\baas#1#2#3{{Bull. Am. Astron. Soc.}
{\bf #1},~#3~(#2)}
\def\ibid#1#2#3{{\it ibid. }{\bf #1},~#3~(#2)}
\def\cpc#1#2#3{{Comput. Phys. Commun.}
{\bf #1},~#3~(#2)}
\def\astp#1#2#3{{Astropart. Phys.}
{\bf #1},~#3~(#2)}
\def\epjc#1#2#3{{Eur. Phys. J. C}
{\bf #1},~#3~(#2)}
\def\nima#1#2#3{{Nucl. Instrum. Meth. A}
{\bf #1},~#3~(#2)}
\def\jhep#1#2#3{{J. High Energy Phys.}
{\bf #1},~#3~(#2)}

\end{document}